\documentclass[aps,prl,reprint,showpacs,superscriptaddress]{revtex4-1}
\usepackage{graphicx}
\usepackage{dcolumn}
\usepackage{bm}
\begin{document}


\title{The generation of unexpected super-high energetic electrons at relativistic circularly polarized laser-solid interactions in the presence of large scale pre-plasmas}

\author{D. Wu}
\affiliation{Shanghai Institute of Optics and Fine Mechanics, 
Chinese Academy of Sciences, Shanghai 201800, China}
\author{S. I. Krasheninnikov}
\affiliation{University of California-San Diego, La Jolla, California, 92093, USA}
\author{S. X. Luan}
\affiliation{Shanghai Institute of Optics and Fine Mechanics, 
Chinese Academy of Sciences, Shanghai 201800, China}
\author{W. Yu}
\affiliation{Shanghai Institute of Optics and Fine Mechanics, 
Chinese Academy of Sciences, Shanghai 201800, China}

\date{\today}

\begin{abstract}
As an extension of the previous work [arXiv: 1512.02411], we have investigated the role of circularly polarized (CP) laser pulses while keeping other conditions the same. It is found that in the presence of large scale pre-formed plasmas, super-high energetic electrons can be generated at relativistic CP laser-solid interactions. For laser of intensity $10^{20}\ \text{W}/\text{cm}^2$ and pre-plasma scale-length $10\ \mu\text{m}$, the cut-off energy of electron by CP laser is $120\ \text{MeV}$ compared with $100\ \text{MeV}$ in the case of linearly polarized (LP) laser. The unexpected super-high energetic electron acceleration can also be explained by the two-stage acceleration model, {by taking into account the envelop modulation effects of the reflected CP laser pulse.} The underlying physics of this envelop modulation is figured out, and a modified first-stage electron acceleration scaling law in the presence of the modulated-CP laser is also obtained.
\end{abstract}


\maketitle

\section{Introduction}
In our previous work\cite{arXiv}, we have proposed a two-stage electron acceleration model to identify the sources of super-high energetic electrons at relativistic laser pre-formed plasma interactions. The first stage is the synergetic acceleration by longitudinal charge separation electric field and pondermotive forces of reflected laser pulses. A scaling law of the maximal possible electron energy by the first stage acceleration is obtained by solving the motions of electrons in the presence of two cross-propagating laser pulses and an external electric field with limited extension. The fast electrons pre-accelerated by the first stage could expand freely, building an intense electrostatic potential barrier with the potential energy several times as large of the electron kinetic energy. Some of the electrons could be reflected by this potential barrier, with the finial electron kinetic cut-off energy several times higher than their initial values.  

In this work, we have studied the role of circularly polarized (CP) laser pulses at relativistic laser-solid interaction in the presence of large-scale plasmas. There is definitely no doubt that without pre-formed plasmas in front of the solid target, CP laser-solid interaction is a hot-electron-free process, because of the absence of $\textbf{J}\times\textbf{B}$ effect. As an extension of our previous work, we focus on the phenomena of CP laser-plasma interaction at relativistic intensity and in the presence of large-scale pre-plasmas. The computer simulation method is exactly the same as previously\cite{arXiv}, only changing the incident laser pulse from linearly polarized (LP) to CP and keeping other parameters the same. It is found that super-high energetic electrons can be generated by CP laser. For laser of intensity $10^{20}\ \text{W}/\text{cm}^2$ and pre-plasma scale-length $10\ \mu\text{m}$, the cut-off kinetic energy of accelerated electrons by CP laser is $120\ \text{MeV}$ compared with $100\ \text{MeV}$ in the case of LP laser\cite{arXiv}. 

The source of these unexpected super-high energetic electrons can also be explained by the two-stage electron acceleration model\cite{arXiv}. {We have found that the envelop of the reflected CP laser becomes periodically modulated. Here we call it modulated-CP laser instead of CP laser and EP laser, because both its ``polarization direction'' and amplitude (or laser envelop) vary with time.} For the pure CP laser, whose ponderomotive force is zero, the synergetic acceleration in the first stage is of no effect at all. However the envelop of the reflected wave is modulated, and this is enough to turn on the mechanism we are considering. 
A modified scaling law related to the first stage acceleration is obtained, showing that at certain circumstance, CP laser can be more efficient than LP laser at super-high energetic electron generation. The underlying physics of this envelop modulation is also addressed and figured out.

Relativistic laser-solid interaction at pico-second duration in the presence of large-scale pre-formed plasmas is of significant effects on many applications\cite{PhyPla.15.056304, PhysRevLett.104.055002, PhysRevLett.108.115004}, such as fast ignition and bright x-ray sources, et al. Understanding the source of fast electrons is of great importance to these related researches\cite{arXiv,PhyPla.19.060703,PhyPla.21.104510}. 

\section{Simulation results} 
Simulation method and parameters are exactly the same as our previous work\cite{arXiv}, but changing the incident laser from LP to CP. The laser intensity is kept the same $10^{20}\ \text{W}/\text{cm}^2$ with laser wavelength $1\ \mu\text{m}$. The initial plasma is loaded as $n_e=n_{\text{solid}}/(1+\exp[-2(z-z_0)/L_p])$, where $n_{\text{solid}}=50n_c$ is the solid plasma density, $L_p$ is the pre-plasma scale-length, $z_0=180\ \mu\text{m}$ and the size of simulation box is $400\ \mu\text{m}$. Just as we have done previously, we have placed two diagnostic planes to temporally record the electrons passing through to analyse the electron energy spectra. The first one is located at $z=100\ \mu\text{m}$ which records electrons passing through at -z-direction, and the other one is at $z=300\ \mu\text{m}$ recording electrons of z-direction passed.

\begin{figure}
\includegraphics[width=8.00cm]{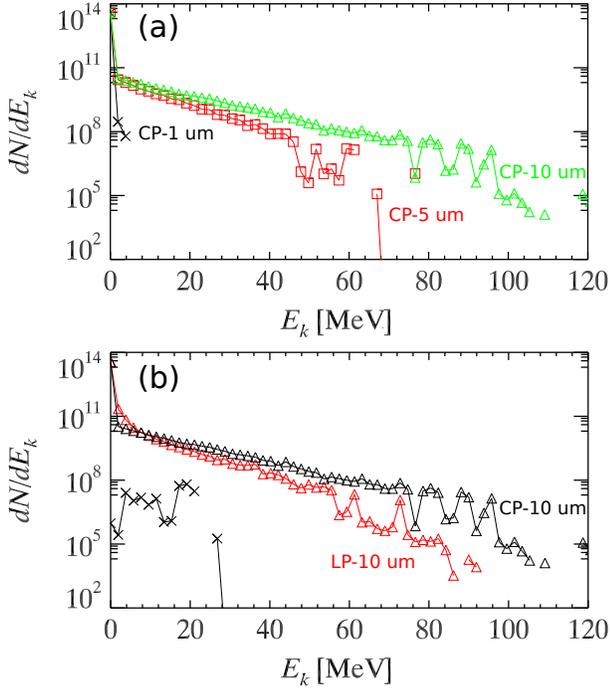}
\caption{\label{fig1} (color online) (a) The final electron energy spectra by CP laser recorded at $z=300\ \mu\text{m}$. Black, red and green lines record the energy spectra for pre-plasma of scale-length $1\ \mu\text{m}$, $5\ \mu\text{m}$ and $10\ \mu\text{m}$, respectively. (b) The black crosses record the spectra of electrons by CP laser collected at $z=100\ \mu\text{m}$ and the black triangles record the spectra collected at $z=300\ \mu\text{m}$ for pre-plasma of scale-length $10\ \mu\text{m}$. The red triangles record the corresponding spectra by LP laser collected at $z=300\ \mu\text{m}$ of pre-plasma scale-length $10\ \mu\text{m}$.}
\end{figure}

\begin{figure}
\includegraphics[width=8.00cm]{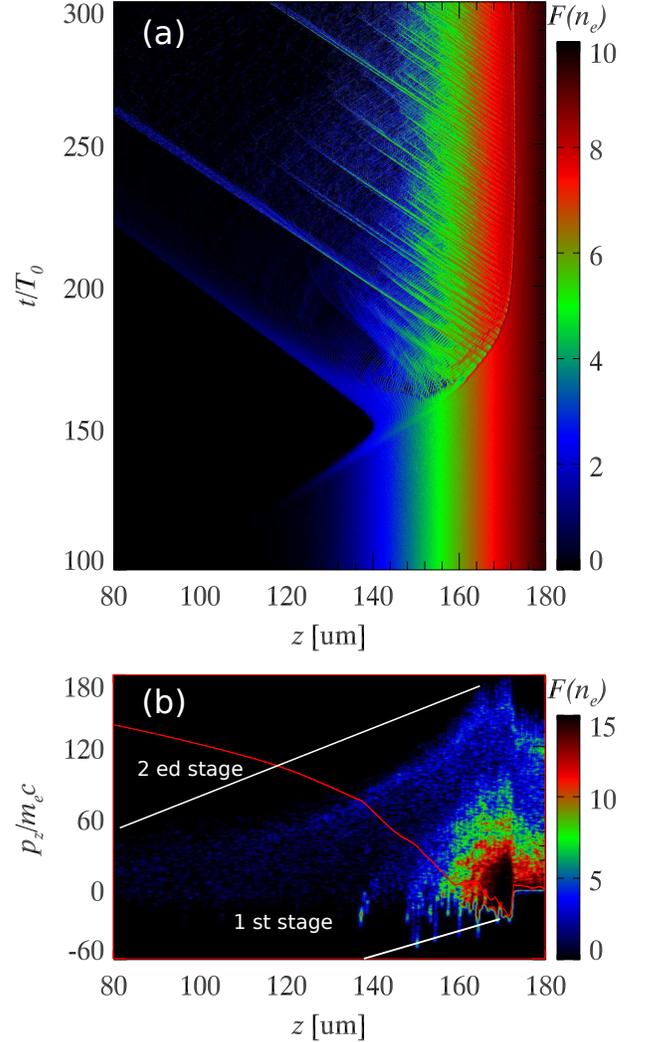}
\caption{\label{fig2} (color online) (a) Space-time evolution of electron density in the case of CP laser with pre-plasma scale-length $10\ \mu\text{m}$ and (b) The corresponding $z$-$v_z$ phase space plot of electrons at $t=300T_0$. The red curve covered on the phase plot is the electrostatic potential.}
\end{figure}   

Fig.\ \ref{fig1} (a) shows the electron energy spectra by CP lasers collected at $z=300\ \mu\text{m}$, indicating that with the increase of pre-formed plasma scale-length, the electron acceleration efficiency is also increasing, which shares the same trend as that of LP lasers. The black lines in Fig.\ \ref{fig1} (b) shows that the cut-off electron energy collected at $z=100\ \mu\text{m}$ is 
$30\ \text{MeV}$ while that recorded at $z=300\ \mu\text{m}$ is $120\ \text{MeV}$, which also shares the same behaviour as that of LP lasers. Compared with the red triangles and black triangles in Fig.\ \ref{fig1} (b), we found that although the $\textbf{J}\times\textbf{B}$ signals are significantly suppressed by CP lasers, the cut-off electron energy by CP laser is obviously higher than that by LP laser. 

Fig.\ \ref{fig2} (a) shows space-time evolution of the electron density driven by CP laser. It is obvious that electrons are firstly accelerated backward (-z direction) and then pulled forward entering into the solid target. From Fig.\ \ref{fig2} (a), we can also find that the backward emissions of the electrons include both isolated bunches and continuous beams. As we have analysed previously\cite{arXiv}, both multi-isolated-electron-bunches and continuous electron beams can build up intense electrostatic potential barrier with its value several times as large of the electron initial kinetic energy. When some of the electron are reflected by the large potential barrier and return back, their final kinetic energies will be several times as large of their initial values. The two-stage acceleration process is also clearly confirmed by Fig.\ \ref{fig2} (b).  

\begin{figure}
\includegraphics[width=7.50cm]{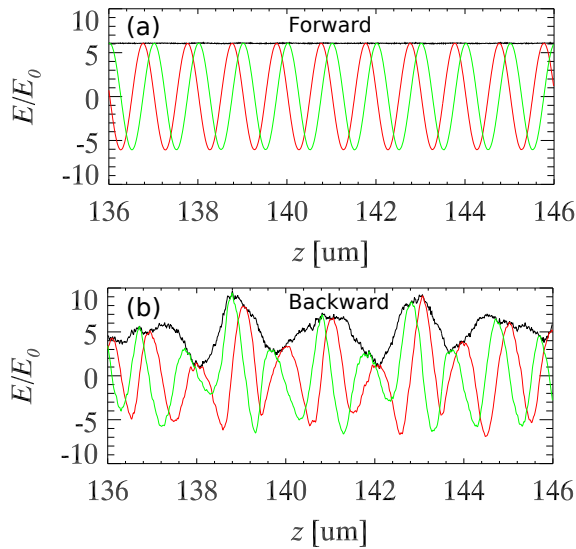}
\caption{\label{fig3} (color online) (a) The z-propagating laser electric field at $t=300T_0$ and (b) the corresponding -z-propagating laser electric field. Black line is the amplitude envelope. Red and green lines are the corresponding $x$ and $y$ components.}
\end{figure}

\section{Envelop modulation of CP laser} 
It is well known that the ponderomotive force of plane CP laser is zero. If the reflected laser pulse is of CP, the first stage acceleration of our two-stage acceleration model is of no effect. Fig.\ \ref{fig3} (a) and (b) show the z-propagating laser electric fields and -z-propagating electric fields. We found that although the incident laser is CP, the reflected one becomes envelop modulated, {where the wave number (frequency) of the modulation is $\sim k_0 /2$ ($\sim \omega_0/2$) and modulation depth is $\sim 0.5$. We thus call this kind of pulse envelop modulated-CP laser.}

\begin{figure}
\includegraphics[width=7.50cm]{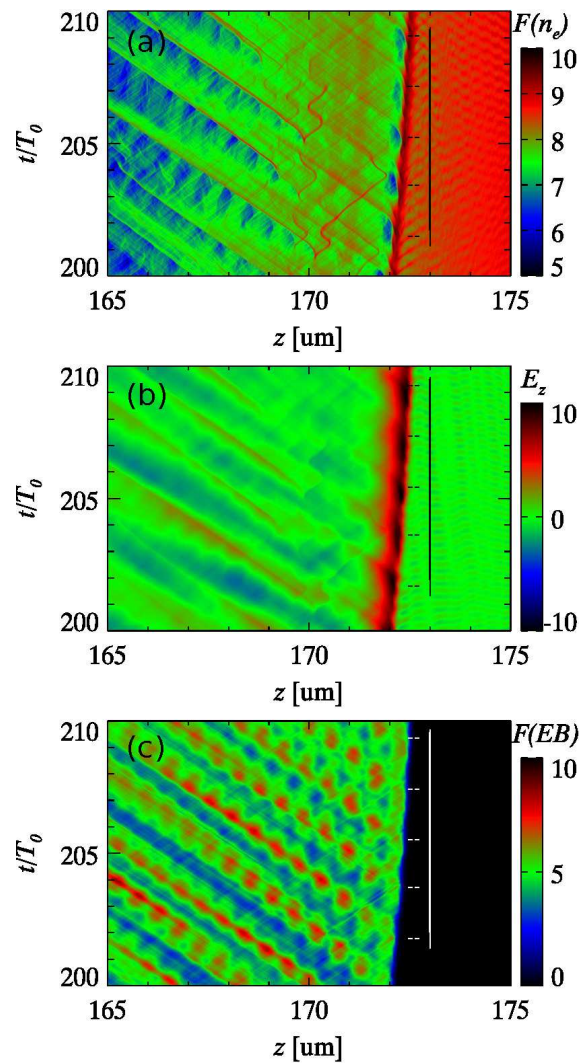}
\caption{\label{fig4} (color online) (a) (b) and (c) space-time evolution of electron density, $E_z$ and laser intensity located at the reflection point, $165\ \mu\text{m}<z<175\ \mu\text{m}$ and $200T_0<t<210T_0$, in the case of CP laser with pre-plasma scale-length $10\ \mu\text{m}$.}
\end{figure}

{To further understand the source of this envelop modulation, we go to the detailed space-time evolution located at the reflection point, $165\ \mu\text{m}<z<175\ \mu\text{m}$ and $200T_0<t<210T_0$. As shown in Fig.\ \ref{fig4}, we find that the oscillation periods of electron density, charge separation field and electromagnetic field energy are also at $2T_0$, and they keep in step with each other.}

{At the reflection point, the laser ponderomotive force and electron charge separation field should balance with each other. This is the case for direct CP laser-solid interactions. However, when there exists large-scale pre-plasmas in front of the solid target, the balance condition will vary with time, when electron leaves the reflection point, the charge separation field therein is lowered and when electrons return back to the reflection point, the charge separation field therein will increase. Actually, the oscillation period of the reflection point can be attributed to the circulating time of electrons in the two stage acceleration process. The dominant oscillation period is determined by the circulating time of the majority of electrons, 
which is $\omega_{pe}$ of the emitting electron beam (Appendix A). 
Due to the ponderomotive force, a sharp density gradient is built at the reflection point, 
with the density of the reflection wall $\gamma n_c$. 
{An energy flux balance condition $\eta I=n_b v_b T_b$ (with $v_b=c$ at ultra high intensities and $\eta\sim10\%$) may then be used to
estimate the ``initial'' density of electron beams, which usually is on the order of $n_c$\cite{Rev.M.Phys.85.751}, 
consistently with the argument that $n_b$ can not exceed the density of the reflection point where hot electrons are generated.}
As electron mass is increase to $\sim\gamma m_e$ ($\gamma=6.0$ for laser of intensity $10^{20}\ \text{W}/\text{cm}^2$ with wavelength $1.0\ \mu\text{m}$), the $\omega_{pe}$ can be calculated to be $\sim1/\gamma^{1/2}\omega_0$, and the oscillation period is 
$\sim\gamma^{1/2}T_0\sim2.4T_0$.}

{Moreover, the modulational (M) and Raman (RS) instability, may also play role in this process. 
It was shown that\cite{PhyPla.2.2807,PhyPla.4.3358,PhyPla.8.3434}, for the relativistic CP laser radiation with $a_0\gg1$ and homogeneous density close to relativistic critical, one cannot distinguish these (M and RS) different modes, and the wave numbers of unstable modes have a broad range. 
While in our case, plasma is inhomogeneous rather strongly, so that if the existing conclusions match our cases need another further study.}

The oscillation of the reflection point will modulate the reflected CP laser. This kind of modulation can also take place in CP laser driven electrostatic shock acceleration of ions\cite{PhyPla.20.023102, PRE.90.023101}, 
and they share the same physics in this work (Appendix B). 
As $a$ is a Lorenz invariant, 
$E=-\partial a/\partial t\sim w a$, $w=\omega_0+\omega_0v_{\text{rp}(t)}$ and $v_{\text{rp}}$ is the velocity of the reflection point, it is easy to see that the oscillation period of the reflection point should be equal to the modulation period of the reflected CP laser. 

\begin{figure}
\includegraphics[width=7.50cm]{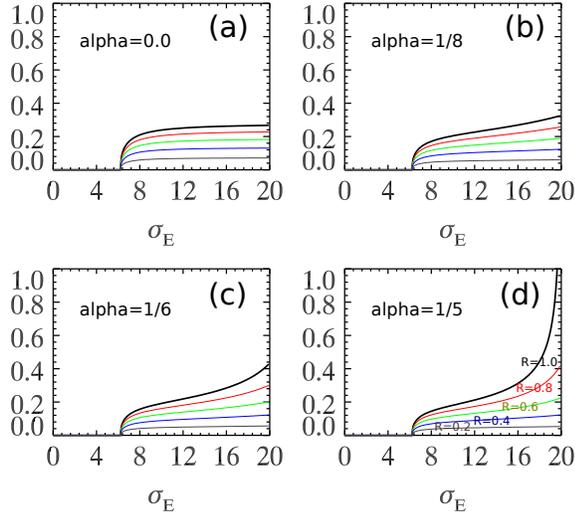}
\caption{\label{fig5} (color online) Coefficient $\eta$ as function of $\delta_E$, reflection ratio $R$ and polarization ratio of EP laser $\alpha$.}
\end{figure}

\section{Modified scaling law} 
The mechanism we proposed can still work under the fact that reflected CP laser is envelop modulated.
Following the method we have developed previously\cite{arXiv} and considering 
$\gamma_A^2=1+a^2/2+\alpha a_{+}^2+(1-2\alpha)a_{+}^2\sin^2(\tau/4)$ with $0<\alpha<1/2$ determining the ``polarization ratio'' of the reflected modulated-CP pulse, we can find the maximal-possible energy gain within the limited longitudinal scale length $L$ and the maximal in-phase time $\tau=2\pi$, which is picked up from Fig.\ \ref{fig3} (b),
\begin{equation}
\label{L}
L=\frac{1}{2E_z^2}[\frac{\gamma_A^2(2\pi+\tau_0)}{\sigma_{E}-2\pi}-\frac{\gamma_A^2(\tau_0)}{\sigma_E}
  -\frac{a_{+}^2}{4}f(\sigma_E)]-{\pi}, \nonumber
\end{equation}
\begin{equation}
\label{Energy}
\Delta \varepsilon(2\pi)=\frac{a_{+}^2}{8E_z}(1-2\alpha)f(\sigma_E),
\end{equation}
where $\sigma_E=\sigma_{\tau_0}/E_z \geq 2\pi$, $a_{+}^2=Ra^2$ with $R$ the reflection ratio, and
\begin{equation}
\label{f}
f(\sigma_E)=\int_{0}^{2\pi}\frac{\sin{(x/2)}}{\sigma_E-x}dx.
\end{equation}

The maximal-possible energy gain within the limited longitudinal length $L$ can be found, by assuming $a\gg1$ and $L\gg1$, 
\begin{equation}
\Delta \varepsilon=\frac{R}{8}(1-2\alpha)\frac{f(\sigma_E)}{g(\sigma_E)} a L^{1/2}=\eta a L^{1/2},
\end{equation}
with $g^2(\sigma_E)=[(R-2\alpha R)\sigma_E+\pi(2\alpha R+1)]/[2\sigma_E(\sigma_E-2\pi)]-{Rf(\sigma_E)}/{2}$. The coefficient $\eta$ is a function of $\tau_E$, $R$ and $\alpha$. As shown in Fig.\ \ref{fig5}, at certain circumstances, the coefficient $\eta$ can be larger than that of LP cases, which is $0.5$. The second stage is the same as that of LP cases.

\section{Conclusions} 
The unexpected super-high energetic electron acceleration at CP laser-solid interaction in the presence of large scale pre-formed plasmas is uncovered by the two-stage acceleration model. A modified first stage scaling law is obtained by including the envelop modulation effects of the reflected CP laser pulse. The underlying physics of this envelop modulation is figured out, and a modified first-stage electron acceleration scaling law in the presence of the modulated-CP laser is also obtained.
The second stage is the same as that of LP cases.

\begin{acknowledgments}
This work was supported by the National Natural Science Foundation of China (11304331, 11174303, 61221064), the National Basic Research Program of China (2013CBA01504, 2011CB808104) and USDOE Grant DENA0001858 at UCSD.
\end{acknowledgments}

\begin{appendix}

\section{Appendix A: On the oscillation of reflection point}

\begin{figure}
\includegraphics[width=8.50cm]{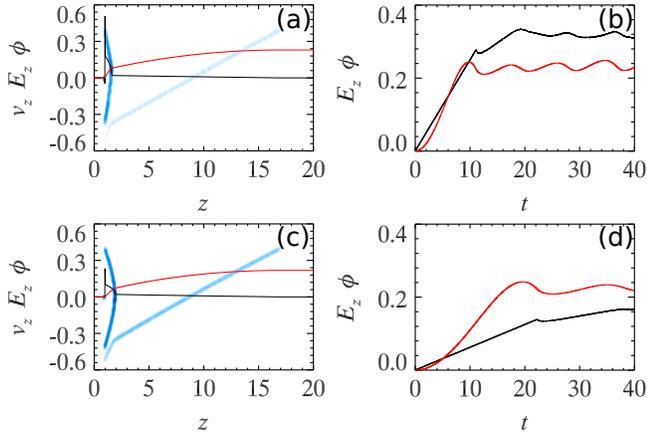}
\caption{\label{fig6} (color online) An uniform electron beam with constant velocity $v_0=0.4$ and different density is emitted, where (a) and (b) correspond to $\omega_{pe}=0.25$, and (c) and (d) correspond to $\omega_{pe}=0.125$. $E_z$ in back line and $\phi$ in red line.}
\end{figure}

\begin{figure}
\includegraphics[width=7.50cm]{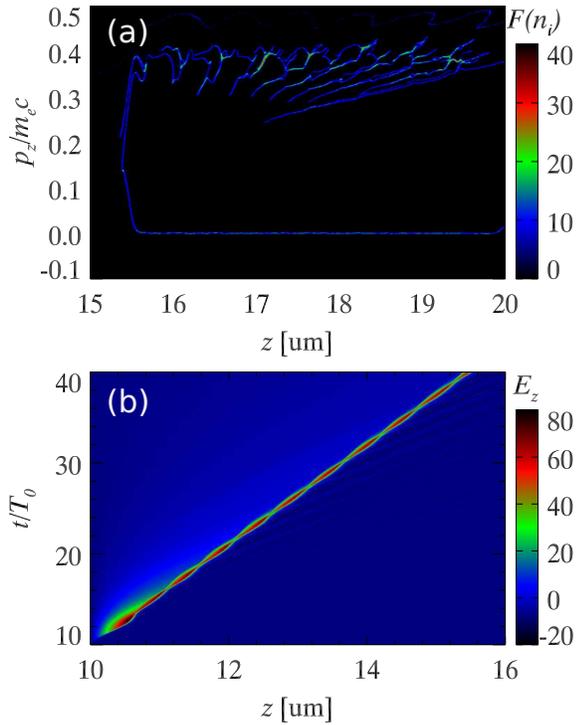}
\caption{\label{fig7} (color online) (a) The $z$-$p_z$ phase plot of ions. (b) the space-time evolution of charge separation electric field.}
\end{figure}

Electron beams of constant density is emitted. 
When the density of the electron beam is 
changed from $n_e=0.25\times0.25$ ($\omega_{pe}=0.25$) to $n_e=0.125\times0.125$ ($\omega_{pe}=0.125$), as shown in Fig.\ \ref{fig6}, the oscillation period of maximal charge separation field and electrostatic potential is increased by twice accordingly.   

\section{Appendix B: CP laser driven shock acceleration}

\begin{figure}
\includegraphics[width=7.50cm]{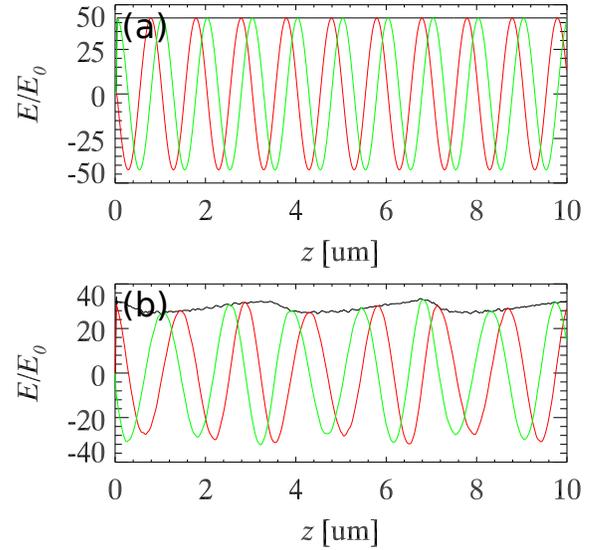}
\caption{\label{fig8} (color online) (a) and (b) are incident CP laser and reflected modulated-CP laser.}
\end{figure}

The CP laser of amplitude $a=60$ propagates into the simulation box from the left boundary. The thick target consists of two species: electrons and protons, which are initially located in the region $10\ \mu\text{m}< x < 20\ \mu\text{m}$ with density $n_e=20n_c$. 
Fig.\ \ref{fig7} (a) shows the $z$-$p_z$ phase plot of the ions. 
The reflect of the ions involves a finite time.
When the ions stream into the shock region, the total charge density grows and a large electrostatic field is induced, 
and vice versa. Fig.\ \ref{fig7} (b) shows the modulated electric field $E_z$, which is on the order of $\omega_{pi}$ \cite{PhyPla.20.023102}. The reflected CP laser is also modulated, as shown in Fig.\ \ref{fig8}, with the modulation period equal to that of $E_z$.

\end{appendix} 

{}

\end{document}